\documentclass[12pt]{iopart} 

\usepackage{enumerate}

\expandafter\let\csname equation*\endcsname\relax
\expandafter\let\csname endequation*\endcsname\relax
\usepackage{amsmath, amsthm, amssymb, amsfonts}

\usepackage{graphicx}
\usepackage[colorlinks=true,linkcolor=blue,citecolor=blue,filecolor=cyan,urlcolor=blue,breaklinks=false]{hyperref}

\usepackage{cite}
\usepackage{color}

\newcommand{\kB}{k_\mathrm{B}}

\newcommand{\mean}[1]{\left\langle #1 \right\rangle}

\begin{document}
\title{Universal bound on the efficiency of molecular motors}
\author{Patrick Pietzonka$^1$, Andre C. Barato$^2$, and Udo Seifert$^1$}

\address{$^1$ II. Institut f\"ur Theoretische Physik, Universit\"at Stuttgart, 70550 Stuttgart, Germany}
\address{$^2$ Max Planck Institute for the Physics of Complex Systems,
  N\"othnitzer Stra\ss e 38, 01187 Dresden, Germany}

\begin{abstract}

The thermodynamic uncertainty relation provides an inequality relating any
mean current, the associated dispersion and the entropy production rate for
arbitrary non-equilibrium steady states.  Applying it here to a general model
of a molecular motor running against an external force or torque, we show that
the thermodynamic efficiency of such motors is universally bounded by an
expression involving only experimentally accessible quantities.  For motors
pulling cargo through a viscous fluid, a universal bound for the corresponding
Stokes efficiency follows as a variant. A similar result holds if mechanical
force is used to synthesize molecules of high chemical potential. Crucially,
no knowledge of the detailed underlying mechano-chemical mechanism is required
for applying these bounds.
\end{abstract}

\vspace{1cm}

\section{Introduction}

Molecular motors are small machines that can transform free energy liberated
in a chemical reaction into mechanical work. Their thermodynamic efficiency
$\eta$ is defined as the ratio between the work exerted against an opposing
mechanical force $f$ (or torque for a rotary motor) and the free energy
consumed in the chemical reaction driving the motor. This efficiency is
universally constrained through the second law by one
\cite{wang98,parm99,parr02,astu02,kolo07,lau07a,boks09,liep08,astu10,gerr10,toya10,toya11,zimm12,golu12,golu12a,kawa14,wago16}.
Molecular motors are isothermal machines for which $\eta\leq 1$ replaces the
Carnot expression applicable to heat engines \cite{seif12}.

A thermodynamic uncertainty relation for non-equilibrium steady states has
recently been obtained in \cite{bara15} (see \cite{ging16} for a rigorous
proof). This relation involves the rate of entropy production, the average of
any fluctuating current and the dispersion associated with this current. It is
expressed as an inequality, which, \textit{inter alia}, establishes the minimal thermodynamic cost of
precision in biochemical processes like enzymatic reactions
\cite{bara15a,bara15c}. Likewise, it can be used to infer unknown enzymatic
schemes from measuring fluctuations of product formation in the spirit of
statistical kinetics \cite{moff14} and thermodynamic inference
\cite{ribe14,alem15}.  Recently, it has been cast in the form of a design
principle for non-equilibrium self-assembly \cite{nguy15}, and it provides an alternative
method to obtain a lower bound on the entropy production, as compared to
the method from \cite{rold10,rold12}. As an important generalization of this uncertainty
relation, bounds on the whole spectrum of fluctuations of such fluctuating
currents have been derived \cite{piet15,ging16,piet16,pole16,ging16a}.

For the simple paradigmatic case of a driven system moving with time $t$ along a coordinate
$x$, the uncertainty relation reads \cite{bara15}
\begin{equation}
  \label{eq:uncrel}
  \sigma\,\frac{D}{v^2}>\kB,
\end{equation}
where $\kB$ is Boltzmann's constant, $\sigma$ the entropy production rate, $v$
the mean velocity, and $D$ the diffusion coefficient
\begin{equation}
  D\equiv \lim_{t\to \infty}\mean{ (\varDelta x(t) - \mean{ \varDelta
      x(t)})^2}/2t,
  \label{eq:Ddef}
\end{equation}
with $\mean{\dots}$ denoting steady state averages throughout.
From a physical perspective, this relation shows that high precision, i.e., a
small dispersion (or uncertainty), comes at the cost of high entropy
production, i.e., dissipation.

In this contribution, we discuss the consequences of this thermodynamic
uncertainty relation for the efficiency of molecular motors. As a main result,
we show that $\eta$ is bounded by
\begin{equation}
\eta\leq \frac{1}{1+v\kB T/Df}, 
\end{equation}
where $v$ is the (mean) velocity of the motor, $D$ its diffusion coefficient, and $T$
is the temperature. The intriguing aspect
of this bound arises from the fact that $v$, $D$, and $f$ are experimentally accessible
quantities. No knowledge of the underlying chemical reactions scheme is necessary
for applying this bound. Moreover, it holds for a huge class
of motor models, arguably essentially for all models that are thermodynamically 
consistent whether based on discrete states or on a continuous potential as 
often used in ratchet models \cite{parm99,parr02,wang02a}. Likewise, it holds
for complexes of motors pulling cooperatively a single cargo particle as,
e.g., investigated in Refs.~\cite{klum05,gros07,muel08,golu12a}.

\section{Thermodynamically consistent motor model}

\begin{figure}
  \centering
  \includegraphics[width=0.8\textwidth]{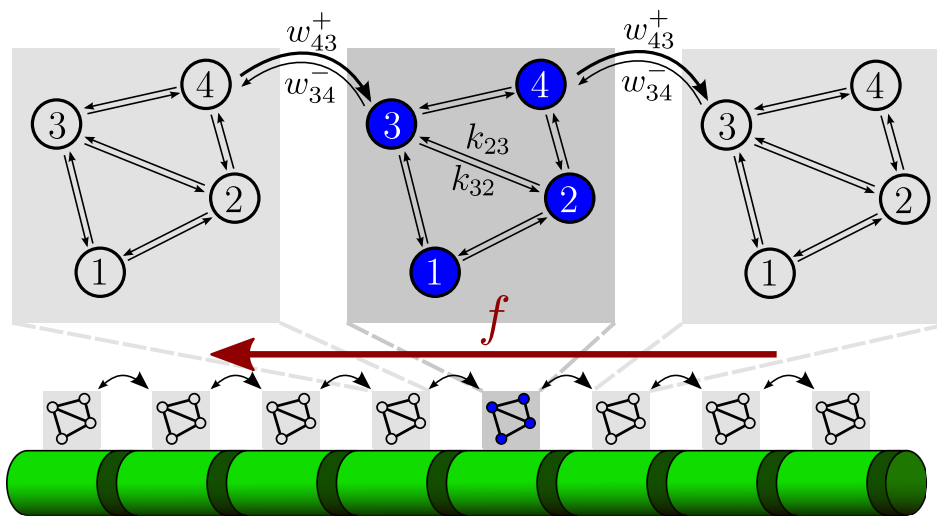}
  \caption{Schematic network of transitions for a molecular
  motor walking along a periodic track against an external force $f$. For each
  periodic interval the internal states $i\in\{1,2,3,4\}$ of the motor can be
  grouped into a ``cell''. }
  \label{fig:mesostates}
\end{figure}

The molecular motor (complex) has an arbitrary number of internal states
$\{i\}$ that describe distinct conformations. Moreover, a possible change in
conformation can be related to the binding of a solute molecule
$A^\alpha$. Here, $\alpha$ label the species, like ATP whose hydrolization to
ADP and $\mathrm{P_i}$ can drive the motor. The motor steps along a periodic
track of periodicity $d$, which means that the set of states $\{i\}$, called a
``cell'', is attached to a discrete linear sequence of spatial positions for
the center of the motor, see Fig. \ref{fig:mesostates}  \cite{piet14}.

Transitions from state $i$ to state $j$ can
occur within a cell, i.e., without net advancement of the motor or can be
accompanied with a forward or a backward step. In the first case, we denote
the rate by $k_{ij}$, in the two latter by $w_{ij}^+$ and $w_{ij}^-$, respectively. The
transitions are microscopically reversible, which means that whenever
$k_{ij}\neq 0$, $k_{ji}$ cannot vanish either. Likewise, $w^+_{ij} \neq 0$
implies $w^-_{ji}\neq 0$ and vice versa. However, there may be transitions
within a cell, which do not occur with spatial motion, i.e., $w^{+,-}_{ij}=0$
is allowed even if $k_{ij}\neq 0$ and vice versa.

Thermodynamic consistency imposes the following two constraints on the rates. First,
for transitions within one periodicity cell, 
\begin{equation}
\frac{k_{ij}}{k_{ji}}= \exp\left[\Big(F_i-F_j+\sum_\alpha b_{ij}^\alpha\mu^\alpha\Big)/\kB T\right].
\label{eq:k}
\end{equation}
Here, $F_{i,j}$ are the free energies of the two states, which may comprise
contributions from the displacement against the external force for incomplete
steps of the motor. If the transition from $i$ to $j$ requires binding a
solute of species $\alpha$ (like, e.g., ATP), then $b_{ij}^\alpha=1$.
Likewise, if this transition leads to a release of such a molecule,
$b_{ij}^\alpha=-1$.  In both cases, the chemical potential $\mu^\alpha$ of
this species enters the expression in the exponent providing a contribution to
the total free energy involved in such a transition. If a transition
additionally involves a step in the forward direction, against the applied
force, then the ratio between such a step and the corresponding backward step
becomes
\begin{equation}
  \frac{w^+_{ij}}{w^-_{ji}}= \exp\left[\Big(F_i-F_j+\sum_\alpha b_{ij}^\alpha\mu^\alpha-fd\Big)/\kB T\right].
\label{eq:w}
\end{equation}

For a fixed applied external force $f$ and externally maintained chemical
potentials $\{\mu^\alpha\}$, the motor reaches a steady state with velocity
\begin{equation}
v=\sum_{ij}p_i(w^+_{ij}-w^-_{ij})d 
\label{eq:v}
\end{equation}
with $p_i$ the steady state probability to find the motor in the internal
state $i$. This velocity
\begin{equation}
v= \mean{\varDelta x(t)}/t = \mean{ n^+(t)-n^-(t)} d/t
\end{equation}
is given by the steady state average of the stochastic displacement $\varDelta x(t)$, where $n^{+,-}(t)$
are the number of forward and backward steps along the track in time
$t$.
In order to apply the thermodynamic uncertainty relation, besides the
diffusion coefficient $D$ of the motor, as defined in Eq.~\eqref{eq:Ddef}, we need the rate of
thermodynamic entropy production $\sigma$. Applying the general principles of
stochastic thermodynamics to this model \cite{seif12}, it is given by
\begin{equation}
\sigma = \kB \sum_{ij}p_i\,\Big(k_{ij}\ln\frac{k_{ij}}{k_{ji}}+w^+_{ij}
\ln \frac {w^+_{ij}}{w^-_{ji}}+w^-_{ij} \ln\frac{w^-_{ij}}{w^+_{ji}}\Big).
\end{equation}
Using the thermodynamic constraints (\ref{eq:k},\ref{eq:w}), the steady state condition 
\begin{equation}
\sum_{j} p_i(k_{ij}+w^+_{ij}+w^-_{ij})=\sum_jp_j(k_{ji}+w^-_{ji}+w^+_{ji}),
\end{equation} 
and the symmetry $b^\alpha_{ij}=-b^\alpha_{ji}$ leads to the expression
\begin{equation}
\sigma = \sum_{ij} p_i(k_{ij}+w^+_{ij}+w^-_{ij})\sum_\alpha b^\alpha_{ij}\mu^\alpha/T-fv/T .
\label{eq:sig}
\end{equation}

The chemical work put into the motor arises from reactions of the type
\begin{equation}
\sum_\alpha r^\alpha_\nu A^\alpha \rightleftharpoons \sum_\alpha s^\alpha_\nu A^\alpha,
\end{equation}
where $r^\alpha_\nu=1$ if species $\alpha$ is an educt
of the forward reaction of type $\nu$ and $s^\alpha_\nu =1$ if species
$\alpha$ is a product of this reaction. In the simplest case of just ATP
hydrolysis, as for F$_{1}$-ATPase \cite{wang98,gerr10,toya11,kawa14} or kinesin
\cite{liep08,fish01,chen02}, we have only one net reaction
\begin{equation}
\mathrm{ATP}\rightleftharpoons \mathrm{ADP} + \mathrm{P_i}.
\end{equation}
In general, the rate of consumption of species $\alpha$ is given by
\begin{equation}
\rho_\alpha = \sum_\nu \gamma_\nu  (r^\alpha_\nu -s^\alpha_\nu),
\end{equation}
where $\gamma_\nu$ is the effective net rate of the reaction of type $\nu$ with $\gamma_\nu<0$
if the reaction goes backward on average. Likewise,
 $\rho_\alpha<0$ if, on average, species $\alpha$ is rather produced than consumed.
 Since
 we assume that all reactions are catalyzed
by the motor acting as an enzyme, thus requiring binding and release as introduced above,
 this net rate can also be written as
\begin{equation}
\rho_\alpha = \sum_{ij}p_i(k_{ij}+w^+_{ij}+w^-_{ij})b_{ij}^\alpha .
\end{equation}
The rate of consumption of chemical (free) energy, i.e., the rate with which
chemical work is put into the motor, becomes
\begin{equation}
\dot w_\mathrm{chem} = \sum_\alpha  \rho_\alpha\mu_\alpha.
\end{equation}  
Inserting these expressions into the entropy production rate \eqref{eq:sig}, we get
\begin{equation}
\sigma = (\dot w_\mathrm{chem}-fv)/T= fv(1/\eta-1)/T 
\label{eq:sigma}
\end{equation}
with the thermodynamic efficiency \cite{parm99}
\begin{equation}
\eta\equiv fv/\dot w_\mathrm{chem}=fv/(fv+T\sigma).
\label{eq:eff1}
\end{equation}

\section{The bound}

Inserting the uncertainty relation \eqref{eq:uncrel} in \eqref{eq:eff1}, we obtain our
main result stated in the introduction, copied here for convenience as
\begin{equation}
\eta\leq \frac{1}{1+v\kB T/Df} .
\label{eq:uni1}
\end{equation}

Remarkably, this universal bound has been derived without any assumptions about the specific molecular mechanism 
driving the motor. It holds for a single motor as well as for complexes of several motors pulling a cargo
to which an external force is applied. Hence, it is sufficient to measure the
velocity and the diffusion constant of the motor to bound its thermodynamic efficiency.
This bound can also be written in terms of the often used randomness parameter $r\equiv 2D/vd$ \cite{svob94,fish01,kolo07,moff14}.
With this quantity, the universal bound on efficiency \eqref{eq:uni1} reads
\begin{equation}
\eta\leq \frac{r}{r+2\kB T/fd}. 
\end{equation}

If only one type of chemical reaction, e.g, ATP hydrolysis is involved, the
rate of chemical work becomes $\dot w_\mathrm{chem}=\rho_\mathrm{ATP} \varDelta \mu$,
where $\varDelta \mu\equiv \mu_\mathrm{ATP}- \mu_\mathrm{ADP}- \mu_\mathrm{P_i}$. The rate of
ATP-consumption can then be bounded as
\begin{equation}
\rho_\mathrm{ATP}\geq \left(f+ \frac{v\kB T}{D}\right)\frac{v}{\varDelta\mu} .
\end{equation}
This relation makes it possible to infer the minimal rate of ATP consumption by
measuring the first and second moments of the displacement of the motor.

\section{Variants}
So far, we have focused on the thermodynamic efficiency $\eta$ where the motor
runs against an external load. For motors pulling cargo through a viscous
environment with no further external load, often the Stokes efficiency
\begin{equation}
\eta_\mathrm{S}\equiv  \gamma^{(0)} v^2/\dot w_\mathrm{chem}
\label{eq:stokesdef}
\end{equation}
is used \cite{wang02a}. It compares the chemical work spent to the (mechanical) power
required by a fictitious force to pull the cargo with a bare friction
coefficient $\gamma^{(0)}$ at the same velocity through the viscous
medium. Using $\sigma=\dot w_\mathrm{chem}/T$ in the uncertainty relation
\eqref{eq:uncrel}, one easily gets
\begin{equation}
\eta_\mathrm{S}\leq \gamma^{(0)} D/\kB  T = D/D^{(0)},
\label{eq:stokesbound}
\end{equation} 
where the second equality follows from the Einstein relation between the bare
friction coefficient and the bare diffusion coefficient $D^{(0)}=\kB
T/\gamma^{(0)} $ of the cargo. Hence the Stokes efficiency is universally
constrained by the ratio between the full diffusion coefficient of the motor
cargo complex and the bare diffusion coefficient of the cargo. Again, this
bound holds independently of all molecular details.

So far, we have assumed that the motor can be described by a set of internal
states replicated along a spatially periodic track using a discrete master
equation dynamics. For motors modeled in a continuous potential possibly
changing due to internal (chemical) transitions \cite{parm99} and for motors pulling big
probe particles described by a Langevin equation \cite{zeld05,zimm12}, these universal results
remain valid. Formally, one has to discretize the spatial coordinate and
introduce (biased) transition rates between neighboring spatial positions.
After such a discretization, which can become arbitrarily fine, one is back at
the model investigated above.

For a motor that performs chemical work $ \dot w_\mathrm{chem}^\mathrm{out}>0$ by
synthesizing molecules with high chemical potential driven by an applied
external force $f^\mathrm{dr}>0$ \cite{toya15} similar results hold with
velocity $v>0$. The efficiency for this situation reads 
\begin{equation}
\eta\equiv \dot w_\mathrm{chem}^\mathrm{out}/f^\mathrm{dr}v= (f^\mathrm{dr}v-T\sigma)/f^\mathrm{dr}v\geq 0.
\end{equation}
The inequality \eqref{eq:uncrel} leads to
\begin{equation}
\eta\leq 1-v\kB T/f^\mathrm{dr}D.
\end{equation}
Again, the efficiency of this molecular machine can be bounded without
measuring the rate at which it produces molecules.

\section{Numerical case study}

\begin{figure}
  \centering
  (a)\raisebox{-\height}{\includegraphics[width=0.55\textwidth]{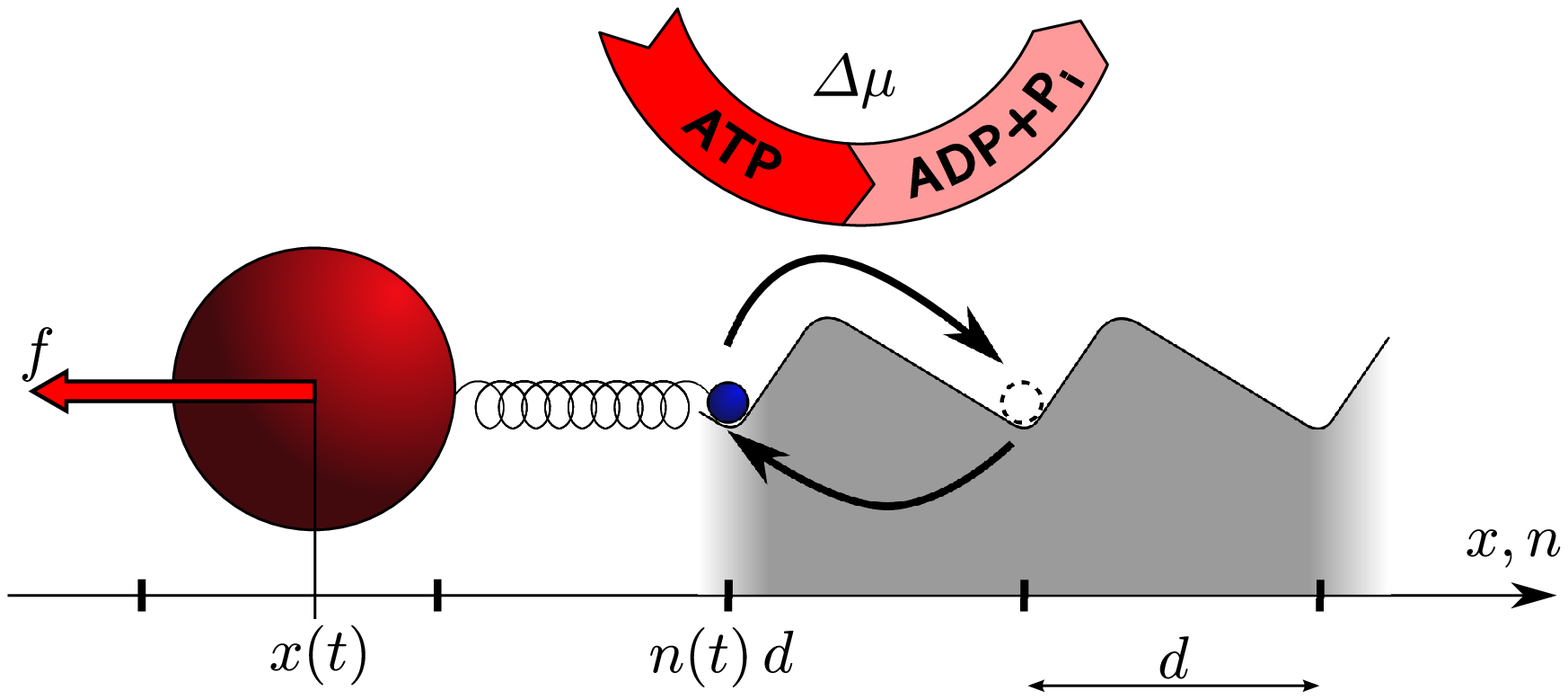}}
  (b)\raisebox{-\height}{\includegraphics[width=0.45\textwidth]{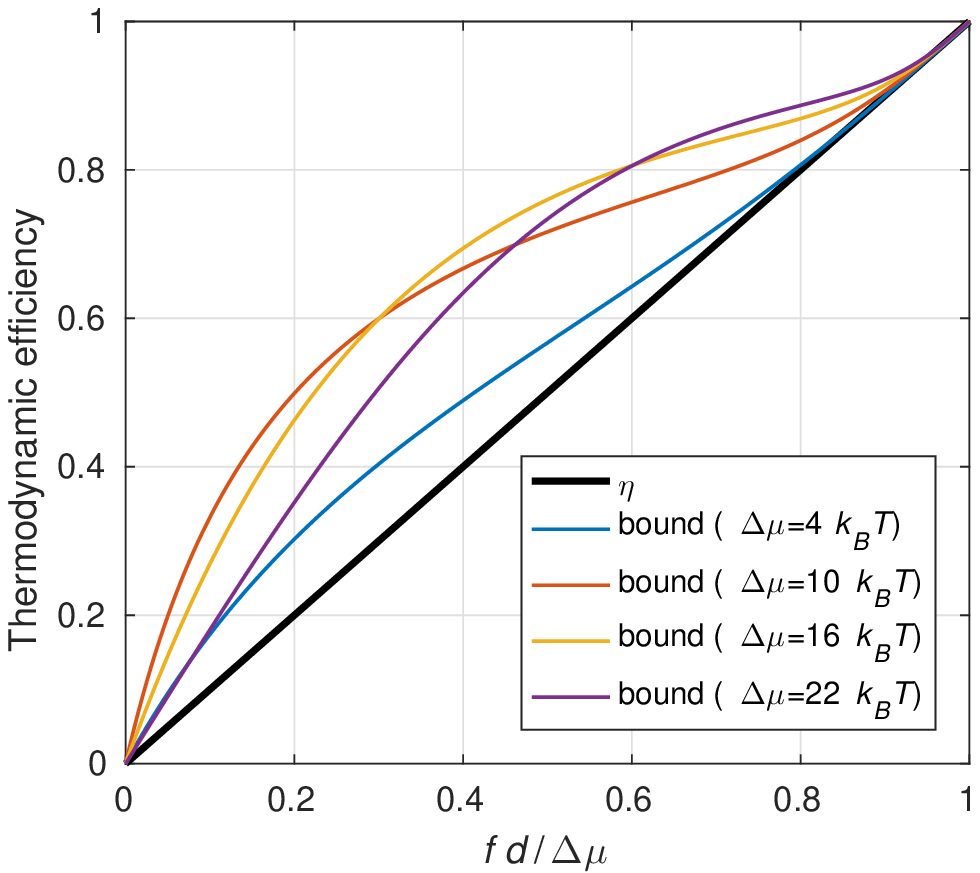}}
  (c)\raisebox{-\height}{\includegraphics[width=0.45\textwidth]{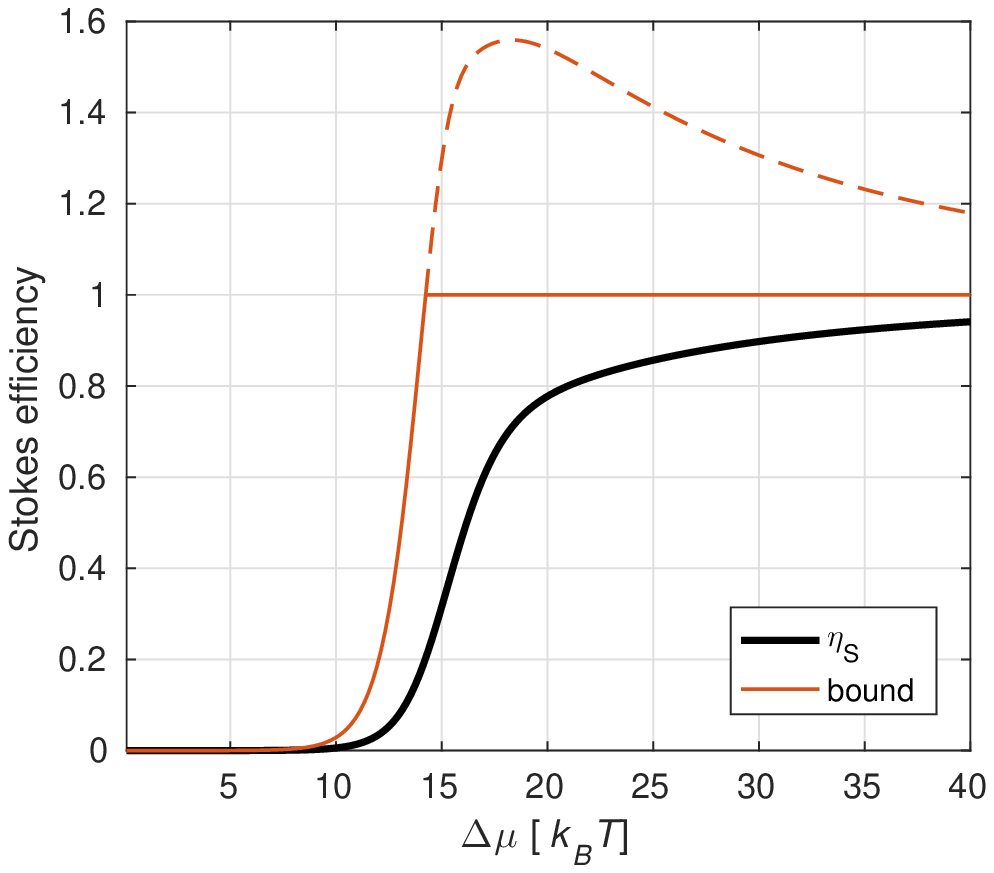}}
  \caption{(a) Illustration of the model for an ATP driven motor (blue)
    elastically coupled to a probe particle (red), to which the external force
    $f$ is applied. (b) Comparison of the thermodynamic efficiency $\eta$ as a
    function of the scaled force $fd/\varDelta\mu$ (black line) and the bound
    \eqref{eq:uni1} at various values of $\varDelta\mu$ (colored curves). (c)
    Comparison of the Stokes efficiency as a function of $\varDelta\mu$
    (black) and the bound \eqref{eq:stokesbound} (red). For large
    $\varDelta\mu$, where the bound surpasses $1$, it is continued as a dashed
    line. Model parameters were chosen as detailed in Ref.~\cite{zimm12},
    different values of $\varDelta\mu$ were generated by varying the
    concentration of ATP while keeping the concentrations of ADP and
    $\mathrm{P_i}$ fixed.}
  \label{fig:numerics}
\end{figure}

We illustrate the bound for a simple model of the rotary motor F$_{1}$-ATPase
coupled elastically to a colloidal probe particle \cite{noji97,zimm12}. As
shown in Fig.~\ref{fig:numerics}a, after mapping the rotary motion to a linear
one, the motor performs discrete steps of length $d$, with each forward step
hydrolizing an ATP molecule and each backward step synthesizing an ATP
molecule. The stepping rates depend on the chemical potential difference
$\varDelta\mu$ of the ATP reaction and the elongation of the linker to the
probe particle. The continuous dynamics of the probe particle with friction
coefficient $\gamma^{(0)}$ is modeled as overdamped Brownian motion subject to
the external force $f$ and the potential force of the linker \cite{zimm12}. 

The average velocity $v$ and the diffusion constant $D$ for this model can be
calculated numerically from the master equation that is obtained by finely
discretizing the possible elongations of the linker and truncating at large
elongations. Due to the tight coupling between the chemical reaction and the
motion of the motor, the rate of the chemical work is $\dot
w_\mathrm{chem}=\varDelta\mu v/d$ \cite{zimm12}, so that the thermodynamic efficiency
\eqref{eq:eff1} becomes trivially
\begin{equation}
  \eta=f d/\varDelta\mu
  \label{eq:td_uc}
\end{equation}
while the Stokes efficiency \eqref{eq:stokesdef} is
\begin{equation}
  \eta_\mathrm{S}=\gamma^{(0)}v d/\varDelta\mu.
  \label{eq:stokes_uc}
\end{equation}

In Fig.~\ref{fig:numerics}b, we compare the thermodynamic efficiency
\eqref{eq:td_uc} to the bound \eqref{eq:uni1} for forces ranging from $0$ to
$\varDelta\mu/d$ and selected values of $\varDelta\mu$. The bound is trivially
saturated for $f=0$ and for the case $fd=\varDelta\mu$, where the chemical and
the mechanical force balance each other, leading effectively to equilibrium conditions
with vanishing velocity. Since the uncertainty relation becomes exact within
linear response for unicyclic systems \cite{bara15,ging16}, the bound is also saturated for forces
close to $\varDelta\mu/d$. Similarly, for $\varDelta\mu\to 0$, the bound approaches $\eta$ for all $0<f<\varDelta\mu/d$.

In Fig.~\ref{fig:numerics}c, the Stokes efficiency \eqref{eq:stokes_uc} is
compared to the bound \eqref{eq:stokesbound} for a range of $\varDelta\mu$ and
constant $f=0$. While this bound is also saturated in the linear response
regime for $\varDelta\mu\to 0$, it becomes rather loose for large values of
$\varDelta\mu$. It even surpasses the bound $\eta_\mathrm{S}<1$, proven in
Ref.~\cite{wang02a}, when $D$ becomes larger than the bare diffusion
coefficient $D^{(0)}$ \cite{haya15,shin16}. Obviously, the bound
\eqref{eq:stokesbound} is not useful in this range of parameters.

\section{Conclusion}

We have derived a universal bound on the efficiency of molecular motors that
depends only on the fluctuating displacement of the motor. The sole knowledge
of both the average and the dispersion of this experimentally accessible quantity
yields a bound that is
independent of the underlying mechano-chemical reaction scheme.
This result applies to any nano or micro machine operating in an environment
of fixed temperature. Extensions to devices coupled to two heat baths of
different temperature should be possible and are left for future research.

\section*{References}

\bibliographystyle{utphys}
\bibliography{/home/public/papers-softbio/bibtex/refs}

\providecommand{\href}[2]{#2}\begingroup\raggedright\begin{thebibliography}{10}

\bibitem{wang98}
H.~Wang and G.~Oster, ``Energy transduction in the {F}$_1$ motor of {ATP}
  synthase,'' \href{http://dx.doi.org/10.1038/24409}{{\em Nature} {\bfseries
  396} (1998) 279--282}.

\bibitem{parm99}
A.~Parmeggiani, F.~J\"ulicher, A.~Ajdari, and J.~Prost, ``Energy transduction
  of isothermal ratchets: Generic aspects and specific examples close to and
  far from equilibrium,''
  \href{http://dx.doi.org/10.1103/PhysRevE.60.2127}{{\em Phys.\ Rev.\ E}
  {\bfseries 60} (1999) 2127}.

\bibitem{parr02}
J.~M.~R. Parrondo and B.~J.~D. Cisneros, ``Energetics of {B}rownian motors: a
  review,'' \href{http://dx.doi.org/10.1007/s003390201332}{{\em Applied Physics
  A} {\bfseries 75} (2002) 179}.

\bibitem{astu02}
R.~D. Astumian and P.~H\"anggi, ``{B}rownian motors,''
  \href{http://dx.doi.org/10.1063/1.1535005}{{\em Physics Today} {\bfseries
  55} (2002) 33}.

\bibitem{kolo07}
A.~B. Kolomeisky and M.~E. Fisher, ``Molecular motors: A theorist's
  perspective,''
  \href{http://dx.doi.org/10.1146/annurev.physchem.58.032806.104532}{{\em Ann.
  Rev. Phys. Chem.} {\bfseries 58} (2007) 675--695}.

\bibitem{lau07a}
A.~W.~C. Lau, D.~Lacoste, and K.~Mallick, ``Non-equilibrium fluctuations and
  mechanochemical couplings of a molecular motor,''
  \href{http://dx.doi.org/10.1103/PhysRevLett.99.158102}{{\em Phys.\ Rev.\
  Lett.} {\bfseries 99} (2007) 158102}.

\bibitem{boks09}
E.~Boksenbojm and B.~Wynants, ``The entropy and efficiency of a molecular motor
  model,'' \href{http://dx.doi.org/10.1088/1751-8113/42/44/445003}{{\em J.
  Phys. A: Math. Theor.} {\bfseries 42} (2009) 445003}.

\bibitem{liep08}
R.~Lipowsky and S.~Liepelt, ``Chemomechanical coupling of molecular motors:
  Thermodynamics, network representations, and balance conditions,''
  \href{http://dx.doi.org/10.1007/s10955-007-9425-7}{{\em J.\ Stat.\ Phys.}
  {\bfseries 130} (2008) 39--67}.

\bibitem{astu10}
R.~D. Astumian, ``Thermodynamics and kinetics of molecular motors,''
  \href{http://dx.doi.org/10.1016/j.bpj.2010.02.040}{{\em Biophys. J.}
  {\bfseries 98} (2010) 2401--2409}.

\bibitem{gerr10}
E.~Gerritsma and P.~Gaspard, ``Chemomechanical coupling and stochastic
  thermodynamics of the {F}$_1$-{ATP}ase molecuar motor with an applied
  external torque,'' \href{http://dx.doi.org/10.1142/S1793048010001214}{{\em
  Biophys. Rev. Lett.} {\bfseries 5} (2010) 163--208}.

\bibitem{toya10}
S.~Toyabe, T.~Okamoto, T.~Watanabe-Nakayama, H.~Taketani, S.~Kudo, and
  E.~Muneyuki, ``Nonequilibrium energetics of a single {F$_1$}-{ATP}ase
  molecule,'' \href{http://dx.doi.org/10.1103/PhysRevLett.104.198103}{{\em
  Phys.\ Rev.\ Lett.} {\bfseries 104} (2010) 198103}.

\bibitem{toya11}
S.~Toyabe, T.~Watanabe-Nakayama, T.~Okamoto, S.~Kudo, and E.~Muneyuki,
  ``Thermodynamic efficiency and mechanochemical coupling of
  {F}$_1$-{ATP}ase,'' \href{http://dx.doi.org/10.1073/pnas.1106787108}{{\em
  Proc.\ Natl.\ Acad.\ Sci.\ U.S.A.} {\bfseries 108} (2011) 17951--17956}.

\bibitem{zimm12}
E.~Zimmermann and U.~Seifert, ``Efficiency of a molecular motor: A generic
  hybrid model applied to the {F}$_1$-{ATP}ase,''
  \href{http://dx.doi.org/10.1088/1367-2630/14/10/103023}{{\em New\ J.\ Phys.}
  {\bfseries 14} (2012) 103023}.

\bibitem{golu12}
N.~Golubeva, A.~Imparato, and L.~Peliti, ``Efficiency of molecular machines
  with continuous phase space,''
  \href{http://dx.doi.org/10.1209/0295-5075/97/60005}{{\em EPL} {\bfseries
  97} (2012) 60005}.

\bibitem{golu12a}
N.~Golubeva and A.~Imparato, ``Efficiency at maximum power of interacting
  molecular machines,''
  \href{http://dx.doi.org/10.1103/PhysRevLett.109.190602}{{\em Phys.\ Rev.\
  Lett.} {\bfseries 109} (2012) 190602}.

\bibitem{kawa14}
K.~Kawaguchi, S.~Sasa, and T.~Sagawa, ``Nonequilibrium dissipation-free
  transport in {F}$_1$-{ATP}ase and the thermodynamic role of asymmetric
  allosterism,'' \href{http://dx.doi.org/10.1016/j.bpj.2014.04.034}{{\em
  Biophys. J.} {\bfseries 106} (2014) 2450--2457}.

\bibitem{wago16}
J.~A. Wagoner and K.~A. Dill, ``Molecular motors: Power strokes outperform
  brownian ratchets,'' \href{http://dx.doi.org/10.1021/acs.jpcb.6b02776}{{\em
  J.\ Phys.\ Chem.\ B} {\bfseries 120} (2016) 6327--6336}.

\bibitem{seif12}
U.~Seifert, ``Stochastic thermodynamics, fluctuation theorems, and molecular
  machines,'' \href{http://dx.doi.org/10.1088/0034-4885/75/12/126001}{{\em Rep.
  Prog. Phys.} {\bfseries 75} (2012) 126001}.

\bibitem{bara15}
A.~C. Barato and U.~Seifert, ``Thermodynamic uncertainty relation for
  biomolecular processes,''
  \href{http://dx.doi.org/10.1103/PhysRevLett.114.158101}{{\em Phys.\ Rev.\
  Lett.} {\bfseries 114} (2015) 158101}.

\bibitem{ging16}
T.~R. Gingrich, J.~M. Horowitz, N.~Perunov, and J.~L. England, ``Dissipation
  bounds all steady-state current fluctuations,''
  \href{http://dx.doi.org/10.1103/PhysRevLett.116.120601}{{\em Phys.\ Rev.\
  Lett.} {\bfseries 116} (2016) 120601}.

\bibitem{bara15a}
A.~C. Barato and U.~Seifert, ``Universal bound on the fano factor in enzyme
  kinetics,'' \href{http://dx.doi.org/10.1021/acs.jpcb.5b01918}{{\em J.\ Phys.\
  Chem.\ B} {\bfseries 119} (2015) 6555--6561}.

\bibitem{bara15c}
A.~C. Barato and U.~Seifert, ``Skewness and kurtosis in statistical kinetics,''
  \href{http://dx.doi.org/10.1103/PhysRevLett.115.188103}{{\em Phys.\ Rev.\
  Lett.} {\bfseries 115} (2015) 188103}.

\bibitem{moff14}
J.~R. Moffitt and C.~Bustamante, ``Extracting signal from noise: kinetic
  mechanisms from a michaelis–menten-like expression for enzymatic
  fluctuations,'' \href{http://dx.doi.org/10.1111/febs.12545}{{\em FEBS
  Journal} {\bfseries 281} (2014) 498--517}.

\bibitem{ribe14}
M.~Ribezzi-Crivellari and F.~Ritort, ``Free-energy inference from partial work
  measurements in small systems,''
  \href{http://dx.doi.org/10.1073/pnas.1320006111}{{\em Proc.\ Natl.\ Acad.\
  Sci.\ U.S.A.} {\bfseries 111} (2014) E3386--E3394}.

\bibitem{alem15}
A.~Alemany, M.~Ribezzi-Crivellari, and F.~Ritort, ``From free energy
  measurements to thermodynamic inference in nonequilibrium small systems,''
  \href{http://dx.doi.org/10.1088/1367-2630/17/7/075009}{{\em New\ J.\ Phys.}
  {\bfseries 17} (2015) 075009}.

\bibitem{nguy15}
M.~Nguyen and S.~Vaikuntanathan, ``Design principles for nonequilibrium
  self-assembly,'' \href{http://dx.doi.org/10.1073/pnas.1609983113}{{\em Proc.
  Natl. Acad. Sci. U.S.A.} {\bfseries 113} (2016) 14231--14236}.

\bibitem{rold10}
E.~Roldan and J.~M.~R. Parrondo, ``Estimating dissipation from single
  stationary trajectories,''
  \href{http://dx.doi.org/10.1103/PhysRevLett.105.150607}{{\em Phys.\ Rev.\
  Lett.} {\bfseries 105} (2010) 150607}.

\bibitem{rold12}
E.~Roldan and J.~M.~R. Parrondo, ``Entropy production and {K}ullback-{L}eibler
  divergence between stationary trajectories of discrete systems,''
  \href{http://dx.doi.org/10.1103/PhysRevE.85.031129}{{\em Phys.\ Rev.\ E}
  {\bfseries 85} (2012) 031129}.

\bibitem{piet15}
P.~Pietzonka, A.~C. Barato, and U.~Seifert, ``Universal bounds on current
  fluctuations,'' \href{http://dx.doi.org/10.1103/PhysRevE.93.052145}{{\em
  Phys.\ Rev.\ E} {\bfseries 93} (2016) 052145}.

\bibitem{piet16}
P.~Pietzonka, A.~C. Barato, and U.~Seifert, ``Affinity- and topology-dependent
  bound on current fluctuations,''
  \href{http://dx.doi.org/10.1088/1751-8113/49/34/34LT01}{{\em J. Phys. A:
  Math. Theor.} {\bfseries 49} (2016) 34LT01}.

\bibitem{pole16}
M.~Polettini, A.~Lazarescu, and M.~Esposito, ``Tightening the uncertainty
  principle for stochastic currents,''
  \href{http://dx.doi.org/10.1103/PhysRevE.94.052104}{{\em Phys. Rev. E}
  {\bfseries 94} (2016) 052104}.

\bibitem{ging16a}
T.~R. Gingrich, G.~M. Rotskoff, and J.~M. Horowitz, ``Inferring microscopic
  dissipation rates from coarse-grained current fluctuations,'' {\em arXiv
  preprint arXiv:1609.07131} (2016).

\bibitem{wang02a}
H.~Wang and G.~F. Oster, ``The {S}tokes efficiency for molecular motors and its
  applications,'' \href{http://dx.doi.org/10.1209/epl/i2002-00385-6}{{\em
  Europhys.\ Lett.} {\bfseries 57} (2002) 134}.

\bibitem{klum05}
S.~Klumpp and R.~Lipowsky, ``Cooperative cargo transport by several molecular
  motors,'' \href{http://dx.doi.org/10.1073/pnas.0507363102}{{\em Proc.\ Natl.\
  Acad.\ Sci.\ U.S.A.} {\bfseries 102} (2005) 17284--17289}.

\bibitem{gros07}
S.~P. Gross, M.~Vershinin, and G.~Shubeita, ``Cargo transport: Two motors are
  sometimes better than one,''
  \href{http://dx.doi.org/10.1016/j.cub.2007.04.025}{{\em Curr. Biol.}
  {\bfseries 17} (2007) R478 -- R486}.

\bibitem{muel08}
M.~J.~I. M\"uller, S.~Klumpp, and R.~Lipowsky, ``Tug-of-war as a cooperative
  mechanism for bidirectional cargo transport by molecular motors,''
  \href{http://dx.doi.org/10.1073/pnas.0706825105}{{\em Proc.\ Natl.\ Acad.\
  Sci.\ U.S.A.} {\bfseries 105}, 4609--4614}.

\bibitem{piet14}
P.~Pietzonka, E.~Zimmermann, and U.~Seifert, ``Fine-structured large deviations
  and the fluctuation theorem: Molecular motors and beyond,''
  \href{http://dx.doi.org/10.1209/0295-5075/107/20002}{{\em EPL} {\bfseries
  107} (2014) 20002}.

\bibitem{fish01}
M.~E. Fisher and A.~B. Kolomeisky, ``Simple mechanochemistry describes the
  dynamics of kinesin molecules,''
  \href{http://dx.doi.org/10.1073/pnas.141080498}{{\em Proc.\ Natl.\ Acad.\
  Sci.\ U.S.A.} {\bfseries 98} (2001) 7748--7753}.

\bibitem{chen02}
Y.~Chen, B.~Yan, and R.~J. Rubin, ``Fluctuations and randomness of movement of
  the bead powered by a single kinesin molecule in a force-clamped motility
  assay: Monte carlo simulations,''
  \href{http://dx.doi.org/10.1016/S0006-3495(02)75250-8}{{\em Biophys. J.}
  {\bfseries 83} (2002) 2360--2369}.

\bibitem{svob94}
K.~Svoboda, P.~P. Mitra, and S.~M. Block, ``Fluctuation analysis of motor
  protein movement and single enzyme kinetics,''
  \href{http://dx.doi.org/10.1073/pnas.91.25.11782}{{\em Proc.\ Natl.\ Acad.\
  Sci.\ U.S.A.} {\bfseries 91} (1994) 11782--11786}.

\bibitem{zeld05}
K.~B. Zeldovich, J.~F. Joanny, and J.~Prost, ``Motor proteins transporting
  cargos,'' \href{http://dx.doi.org/10.1140/epje/i2004-10137-6}{{\em Eur. Phys.
  J. E} {\bfseries 17} (2005) 155--163}.

\bibitem{toya15}
S.~Toyabe and E.~Muneyuki, ``Single molecule thermodynamics of {ATP} synthesis
  by {F}$_1$ {-ATP}ase,''
  \href{http://dx.doi.org/10.1088/1367-2630/17/1/015008}{{\em New\ J.\ Phys.}
  {\bfseries 17} (2015) 015008}.

\bibitem{noji97}
H.~Noji, R.~Yasuda, M.~Yoshida, and K.~Kinosita, ``Direct observation of the
  rotation of {F$_1$}-{ATP}ase,''
  \href{http://dx.doi.org/10.1038/386299a0}{{\em Nature} {\bfseries 386} (1997)
  299--302}.

\bibitem{haya15}
R.~Hayashi, K.~Sasaki, S.~Nakamura, S.~Kudo, Y.~Inoue, H.~Noji, and K.~Hayashi,
  ``Giant acceleration of diffusion observed in a single-molecule experiment on
  ${\mathrm{F}}_{1}\text{-}\mathrm{ATPase}$,''
  \href{http://dx.doi.org/10.1103/PhysRevLett.114.248101}{{\em Phys.\ Rev.\
  Lett.} {\bfseries 114} (2015) 248101}.

\bibitem{shin16}
R.~Shinagawa and K.~Sasaki, ``Enhanced diffusion of molecular motors in the
  presence of adenosine triphosphate and external force,''
  \href{http://dx.doi.org/10.7566/JPSJ.85.064004}{{\em J. Phys. Soc. Jpn.}
  {\bfseries 85} (2016) 064004}.

\end{thebibliography}\endgroup

\end{document}